# Unconventional Universal Computation in Babbage's Analytical Engine

Raúl Rojas
University of Nevada Reno
Department of Mathematics and Statistics

This paper shows that the programming model of Babbage's Analytical Engine, although unconventional, can be harnessed in order to simulate indirect addressing, a capability that was not included in the original instruction set. That is, in a theoretical sense, the Analytical Engine was as universal as computers we have today. We show how to implement indirect addressing for a working memory of fixed size; this makes it possible to simulate a Turing machine with a finite tape. The result is, of course, only of theoretical and historical interest, without any practical implications.

## 1    Introduction

The mathematician Charles Babbage worked on the design of the Analytical Engine (AE) for many years. According to Bromley, the first development phase was mainly conceptual, starting around 1833 and leading to the "basic design" of 1837-1838 [Bromley 1982]. This was also the machine Babbage had in mind when he wrote 26 programs for the AE from 1836 to 1840 [Babbage papers]. Had the AE been built, it would have been the first computer of the world. In a recent article, I described the programming architecture of the AE, distilled from the programs that Babbage wrote for the machine [Rojas 2021]. Babbage returned to redesign the AE around 1856, specially so that it could be finally constructed, but until his death in 1871, only a few mechanisms of the AE had been built.

In this paper, I want to describe the unconventional kind of computation supported by the AE, and how it makes possible to write code that could simulate indirect addressing and several important programming structures that we use today. The AE was idiosyncratic, but it would have been as general-purpose as computers that we use today.

The highlights of the architecture of the AE are:

- There is a clear separation between the mill (processor) and the store (memory),
- The machine used geared wheels in order to represent the digits 0 to 9,
- The AE used ten's complement representation for fixed-point operations,
- Each memory address consisted of a stack of wheels, representing the decimal digits for the different powers of ten, with up to 30 or more digits per memory cell,
- The processor could execute the four basic arithmetical operations and send the result back to the memory unit,
- All operations, addresses for operations, and constants, were stored externally in punched cards,
- The string of operation cards could advance independently from the string of variable (address) cards,
- It was possible to use combinatorial cards to jump forward or backward in the string of operation cards if a numerical condition was met.

Fig. 1 shows a diagram of the programming architecture of the AE. On the left we have the processor, called "mill". On the right we have the memory, called "store" by Babbage. There is a kind of "bus" that can transport numbers from the store to the mill and back. The mill had its own string of operation cards and the store had two double-strings: one double-string for addresses of operands, and one double-string for constants that could be sent to the processor or to memory.

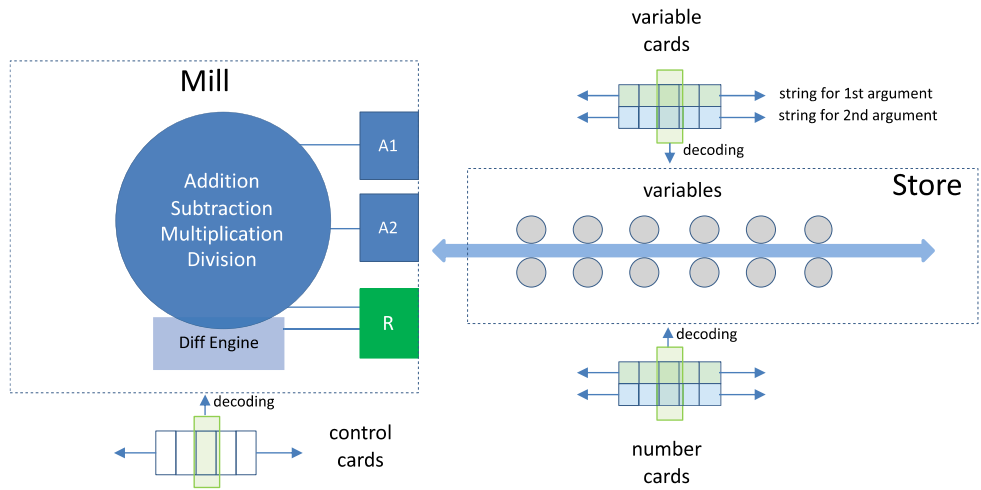

Fig. 1: A diagram of the programming architecture of the AE

When we refer to the "programming architecture", that means that we only refer to the machine from the point of view of the programmer, abstracting from the implementation details, mechanical or electronical. For the AE we have to understand the interplay between the stream of operation and variable cards. The AE was supposed to have a Difference Engine as coprocessor for some calculations, but that is a detail we do not need to refer to here.

## 2      Programming the AE

The programs Babbage wrote for the AE do not represent code in the usual sense, since the decoupling of operations and operands (due to the three separate reading units for punched cards), does not allow us to write something like "add V1,V2,V3" as a primitive operation. In the AE we have two separate and independent strings of cards. For example, if we want to add the contents of addresses 1 and 2, and store it in address 3, we would have the following cards:

| Operation | Addresses | Comments |
|---|---|---|
| + | (V1, 0) | Send first argument to the mill |
|   | (V2, 0) | Send second argument to mill |
|   | (0, V3) | Retrieve result of addition to V3 |

For this operation one single operation card (for addition) requires three pairs of address cards. Also, note that reading operands from memory is a destructive operation, if we want to keep the value of variable 1, we have to send it to a temporary memory cell, and retrieve the value after the arithmetical operation is finished. For example, if that temporary memory cell is the 9th, we would rearrange the stream of punched cards as follows:

| Operation | Addresses | Comments |
|---|---|---|
| + | (V1, V9) | Send first argument to the mill; store temporarily in V9 |
|  | (V2, 0) | Send second argument to mill |
|  | (0, V3) | Retrieve result of addition to V3 |
|  | (V9, V1) | Retrieve V9 to V1 |

This means that an arithmetical operation has a variable number of pairs of cards being read. It is not clear how Babbage indicated to the machine how many pairs of variable cards were used for each operation card, but there should have been some way of doing it.

Essentially, in the AE the memory could send pairs of numbers to be operated upon with addition, subtraction, multiplication and division, and would accept back the result. Since the variable cards only contain fixed address numbers, there is no provision in the AE for using indirect addressing, which is needed in order to access data structures such as vectors and matrices.

In the first published descriptions of the AE, the encoding of the variable cards was not considered in detail [Menabrea 182, Menabrea and Lovelace 1843], but that encoding is used in the programs contained in the Babbage archive [Babbage papers].

## 3      Conditional jump

Babbage did include in the design of the AE the possibility of testing a given memory cell and produce a conditional branch, if the tested value was zero [Bromley 2000]. If the condition is satisfied, a number $n$ is read from the operation cards and a number $m$ from the variable cards. Then, both strings of punched cards are rewound $n$ and $m$ cards, respectively. If the test fails, execution continues with the next instruction and its corresponding cards. This test is used mainly for coding conditional loops, where a piece of code is executed repeatedly, until a counter runs down to zero. The programmer has to keep track of the cards in each stream, because $n$ and $m$ are entered as concrete constants for a jump (it is the distance from the current cards to the continuation cards, after the jump). If we are testing variable 1, we could write this instruction symbolically as

| Operation | Addresses |
|---|---|
| TST | (V1, 0) |
| $n$ | ($m$, 0) |

It is unclear if the $n$ value for the operation cards could be encoded in the same card, together with the instruction opcode, but this is irrelevant. I am assuming that both $n$ and $m$ require a card each.

It is unclear whether the AE could only rewind the strings of punched cards, or also advance the cards forward. Bromley [2000] did not find an explicit indication allowing a jump forward, but also nothing precluding that possibility. The minimal assumption would be that the cards can only be rewound, but I show below how to fix this problem, so that we will provisionally assume that the cards can be advanced forward and backwards. This was in fact possible for the string of constant cards, as program L23 in the digital Babbage Archive shows [Rojas 2021].

## 4     Simple fixed loops

The AE had a provision for repeating an operation a fixed number of times, for example, an addition. The programmer simply encoded the number of repetitions, as shown in the code below (this is also mentioned in [Menabrea and Lovelace 1843]).

| Operation | Addresses |
|---|---|
| 2 + | (V1, 0) |
|  | (V2, 0) |
|  | (0, V3) |
|  | (V4, 0) |
|  | (V5, 0) |
|  | (V6, 0) |

This simple loop, with two repetitions, is adding V1 and V2 and is storing the result in V3, and after that, it is adding V4 and V5 and is storing the result in V6. This example makes clear why Babbage separated the operation cards from the memory cards: operations could be performed repeatedly, acting on a different set of arguments. The absolute addresses had to be given explicitly. Babbage also included the possibility of accumulating a set of memory cells, adding their contest successively in an accumulator, but these are details that we are not relevant here.

## 5    Indirect addressing

From what has been said, it is clear that programming structures such as *if-then-else* and *do-while* could have been implemented using conditional branching. The main problem, for writing modern code, is that indirect addressing is lacking. But this problem can be fixed.

Assume that we write code for the AE using the operations mentioned before and one "subroutine", which corresponds to several basic operations. Let's say that we want to retrieve the contents of a memory cell whose address is contained in variable C. We want to move the contents of that address to address 101, where it will be useful later. Symbolically we want to have something like

$$\text{LOAD } M(C), M(101)$$

That is, load from the memory address C and store it in address 101. We need some auxiliary memory addresses: absolute address Z contains always zero, address E always a one, and addresses T1, T2, etc. are temporary memory cells. They are fixed addresses (numbers).

We can then have at the end of the program a block of instructions referring to the working memory addresses that we want to use. Assume that we only want to use the addresses from 1 to 100. Then the code could be:

| Operation | Addresses | Comments |
|---|---|---|
| − | (C,0), (E,T1), (0,C),(T1,E), | Subtract one from C |
| TST | (C,T1),(T1,C) | Test if counter=0 |
| n1 | m1 | If not, jump to next test |
| + | (V1,T1),(Z,0),(0,V101),(T1,V1) | If yes, move V1 to V101 |
| − | (C,0), (E,T1), (0,C),(T1,E), | Subtract one from C |
| TST | (C,T1),(T1,C) | Test if counter=0 |
| n2 | m2 | If not, jump to next test |
| + | (V2,T1),(Z,0),(0,V101),(T1,V2) | If yes, move V2 to V101 |
| …. | …. | … |

In the code, the move operation is implemented using addition, of the variable to be moved to V101, with zero. As you can see, the code decreases the counter C for the address, and if it reaches zero in the first test, the selected "indexed" memory address is 1, and we move V1 to V101. If not, execution jumps down and we decrease the counter again. If the result is zero, we move V2 to V101. And so on. Since a decreased counter which has reached 0, will never be zero again, execution will just go through the remaining cards, until the end.

Alternatively, once it is clear that we have used the correct memory address, we can force a jump to the end of this section of code by testing the variable Z, which is always zero, and jumping to the end of execution. The necessary cards could be easily inserted in the stream of cards in the table above.

It should be clear that just by jumping to this section of the code we can implement the subroutine "LOAD M(C), M(101)". But we have to return to the point in the code where the subroutine was called. For that, before using the subroutine, a unique number for each jump to this section of code that handles indirect addressing, must be stored in a temporary variable J. At the end of the code, we do something like what we did for selecting indirect addresses. That is, we test successively the value of the variable J and jump back to the position in the code where the call to indirect addressing happened. For that we have to know how many operation and variable cards are between the position of the test and the position the execution has to return to. It would be like having a "GOTO n", where n is the label of the instruction following the "Load" macrooperation. The programmer is responsible for keeping track of the number of cards between instructions.

With that provision we could write something like the following code, where each individual instruction is shown in "high-level code" that the reader can easily translate into the corresponding operation and variable cards. We are using two indirect addresses, so that the number C is loaded twice, once from M(104) and then from M(105). A, B, C, F are memory addresses:

```
            M(104) → C
            L1 → J
            LOAD M(C), M(101)
Label L1:   M(101) → A
            M(105) → C
            L2 → J
            LOAD M(C), M(101)
Label L2:   M(101) → B
            A+B → F
```

With this code, the contents of M(104) and M(105) are added and stored in variable F. The numbers L1 and L2 stored in J identify the position in the code where execution has to return to, and the code managing the jump has to use the adequate number of cards to be skipped, as entered by the programmer keeping track of the design of the program.

## 6      Conditional branching only backwards

If the AE had been only capable of jumping backwards in the strings of cards, we can solve that problem by gluing both ends of the strings of cards, forming a cycle of cards, both for operations and for addresses. Then, even if we can only jump up, we can effectively jump down, in both streams, by jumping up enough cards in the loop of cards so that we return "from below" to the desired code and variable cards section. I am assuming that the number of cards that can be skipped is not bounded by a small constant.

Since Bromley [2000] is of the opinion that the microcode of the AE could have easily allowed both directions for a jump, we could assume that looping the streams of cards would not have been necessary.

## 7    Conclusions

We have shown that there is a way of implementing indirect addressing in the programming model of the AE. The technique is, of course, very inefficient, but it is something theoretically possible for the abstract AE. In fact, I have shown elsewhere that a machine capable of the four arithmetical operations and without conditional branching, but executing a single loop, is, in fact, universal (given enough memory for each given problem, as modern computers require) [Rojas 1996]. The proof relies in an extremely inefficient arithmetical encoding that can simulate conditional branches within a single loop of code, so that the reader should not think that the programming technique discussed in this paper is actually the worst-case in inefficiency. It is in fact a fair compromise, considering the programming architecture of the AE which was not conceived for indirect addressing, if we want to have it in our machine.

It is easy now to envision a program for the AE simulating a Turing machine, with its tape, state and state table, for a finite tape (limited only by the size of the memory of the AE). That shows that the AE would have been capable of handling all kinds of data structures, as a Turing machine can do.